\documentstyle[12pt]{article}
\begin{document}
\begin{center}

{\Large \bf Revealing of the simplest mechanisms of a structure formation}

\vspace{0.5cm}

Olga V. Kirillova
\vspace{0.2cm}

Department of Theoretical Physics, St.Petersburg State University

Ulyanovskaya str. 1, St.Petersburg, 198904 Russia

(e-mail:kirill@snoopy.phys.spbu.ru)


\end{center}

\vspace{1cm}

{\small Abstract.
In this paper, results of investigations of the simplest mechanisms of a 
structure formation are presented. In frameworks of the suggested model
the main attention was focused on such characteristics as wiring of the system, 
clusters formation, dynamics of the wiring. 
The idea to take into account an 
influence of the environment factor is employed 
in the proposed model.
Investigations of systems with such principle of a structure formation
reveal 
that the system's dynamics has typical features of self-organized
criticality 
phenomenon. In the avalanche-like processes, which occur in the wiring 
dynamics, a power law was found with the index close to 1.4.  It
is
independent 
on the environment factor (which in a sense can be considered as system 
parameter). The system wiring is approximated pretty well by the Gaussian 
distribution. The size of the system does not play any role in the dynamics
of the model.}
\vspace{0.5cm}


\section{Introduction}
Nowadays much attention is paid to studying and modeling of different 
systems and its dynamics \cite{mean4bsm}. Usually in such investigations the 
structure of a system is always supposed to be determined and constant, or 
totally random. On the other hand, there is a number of works devoted
exactly 
to studying of structure formation \cite{str} but, as a rule, it has very 
narrow specifics. In the sense of a system evolution, dynamics
undoubtedly 
depends on the geometry of a system \cite{Sn,Nat} and includes
 the structure formation as well. Moreover this property of the system 
dynamics have to be regarded as the fundamental and general for a wide
class of 
phenomena. So the process of the structure formation is one of the most 
important in the world: from physics (atomic structure) to cosmology 
(structure of the Universe), from biology (evolution and morphogenesis) to 
social sciences (formation of different social groups).

The term {\it structure} can be understood in the sense of the spatial 
connection of objects' parts or interactions chains, like in biology:
predator -- prey relationship, foods chains, gene networks. 
In our opinion, the idea seems very attractive that the rise of a
structure
as
a result of natural 
processes taking place within a system occurs in a way common for a wide 
class of phenomena .

As an example of the simplest model of a spatial structure formation, let
us 
take the diffusion-limited aggregation model (DLA) \cite{DLA}. As it was 
pointed out \cite{DLA} such models demonstrate formation of the 
spatial structure with fractal properties. Another attempt to reveal a 
structure, had been done in the investigations \cite{R}
devoted to modeling of networks 
of gene expression. Much attention is paid to this question 
in graph theory \cite{Har}. But all the studies are only particular
quests 
in a narrow field, and moreover today they seem to be absolutely 
disconnected.

From the standpoint of theoretical physics, it is out of the question that 
the topology plays a crucial role in that of what phenomena can take
place in 
a model under study. So it would be interesting to construct a model in 
which a topology (a structure of interaction) arises in the course of 
dynamics. In this work we propose a simplest  model describing structure 
formation in the dynamics of a stochastic system, and report results of 
the study. Let us emphasize that even simplest models reveal interesting 
regularities.

\section{The Model}
Our world is undoubtedly very complex, so one might guess that to gain a 
detailed description it is necessary to write more and more sophisticated 
equations, to include more parameters and variables. On the one hand, it
is true. But on the another, it is important to find the right point of view: 
to reveal fundamental mechanisms of a phenomenon and to check them out on
a model. In the latter case, the simpler (but still nontrivial) model is,
the more valuable it is since we can be sure that we gain a background 
for more detailed description. So our aim is not details but general 
comprehensive knowledge of how complex systems evolve. It is desirable
that 
models would be free of fitting parameters so that one can study a
mechanism 
itself but not parameter-dependent regimes of a model. In our model, besides 
a size of a system we have only one parameter $\alpha$ that can be 
conditionally called {\it the factor of an environment influence}.

Dynamics of the model is following.
At the beginning we have the system which consists of  $N$ nodes 
and  zero links. On the second iteration step a link between a 
pair of elements in the 
system appears with probability $\alpha$ and nothing happens
with probability $1-\alpha$.
Some time later we shall have a number of {\it clusters} (connected 
group of nodes) in the system. (The size
of a cluster is the number of nodes, which forms the cluster.)
At the {\it K}-th iteration step one of the nodes is randomly choosen.
If it is not belong to any clusters, in other words has no 
links, there is possibility of a link appearing (with 
probability $\alpha$) in the system and the situation is not changed with 
probability $1-\alpha$. If the choosen element 
belongs to a cluster, we "check" each link going from the
element and with probability $\alpha$ it is preserved, while it is
 destroyed with probability $1-\alpha$. Thus at the {\it K}-th iteration 
step it is possible a creation of one new link and destroying 
several links. And so on for the {\it K}+1, {\it K}+2, ... iteration steps.
 Let us envisage the next analogy. It is possible to consider our model
as a formation of a network of channels in a medium, which at the 
beginning is
homogeneous (water in rock, electrical current in a medium with high
resistance). By some means we increase pressure/potential in a point of
the medium. If there are not channels going from the point then as the
action result (if pressure/potential exceeded some threshold (it
corresponds to parameter $\alpha$ of the model)) a hole occurs and the
new channel is created.
 If a chosen point already has channels, it seems to be fairly logical
that flow/current will use the existing ones rather than to make a new
hole.

Thus at the beginning we have homogeneous medium, after some time we
get porous one.

One can see, that the system dynamics allows, in principle, formation of a
cluster of any size, up to the size of the system. Obviously the larger a
cluster is, the less stable it is. Let us also point out that in general
case the dynamics of such systems does not have attractors, i.e. it can
be considered as "the evolution with open end".

The proposed model seems to be close to the models investigated in
the
percolation theory \cite{perc,Efr}. Taking into account that
'percolation theory deals with connectivity of a very large
(macroscopic) number of elements under condition that the relation of
every
element with its neighbors has a random, but quite definite character,
(for example, setting up by the means of using of a
random number generator possess definite properties)' \cite{Efr} we might
emphasize that our model neither has any fixed structure of the system
(that has to be understood in the sense of neighbors definition
procedure) nor any determined space dimension in this sense it is similar
just to infinite-range percolation \cite{GF}.
 In the suggested model a structure arises in the
process of system evolution, it is mobile, in this aspect it is 
closer rather to the second order phase transition \cite{Lan}. 
There is the chemistry application of reversible gelation (cooking gelatine)
where similar creation and destruction of bonds occurs. Let us
stress that dynamics of our model essentially differs from \cite{A,2},
we do not consider a disorded lattice as well as particles motions
in
dynamically disorded medium. We do not study a transport task \cite{3}. We
envisage the mobile structure creation processes. But deep in our mind
we are interested in the same problem -- a role of connectivity
in a system.

\section{Results of computer simulations}

As the main characteristics of a forming structure we consider: the
probability distribution of the system wiring (in this context the wiring
is a ratio of the number of links to the number of elements), the average
fraction of free elements (the elements which have no neighbors) in the
system, the distribution of clusters as
a function of size.

We have studied these characteristics for different values of parameter
$\alpha$.

It has been turned out that the probability distribution of the
system
wiring is excellently approximated by the Gaussian law, i.e.
$$
p(x)=\frac{1}{\sigma\sqrt{2 \pi}} e^{\frac{-(x-a)^2}{2\sigma^2}}
$$
where $x$ is the wiring
(Fig. 1)

The distribution is localized on the very narrow interval of the wiring
value. The results for the average and deviations for different $\alpha$
are given in TABLE 1. In (Fig. 2) one can see the distributions
for different values of $\alpha$.

Turning to the distribution of clusters as a function of its size, one
can see that the character of the dependence is exponential and does not
depend on a system size. (Fig. 3). 
$$ n_s \propto e^{\beta s}$$ where $n_s$ is the number of clusters
consisting of $s$ nodes.
For
$1/4<=\alpha<=3/4$ the accuracy of approximation varies from 0.999067
($\alpha=3/4$) to 0.999862 ($\alpha=1/2$) and for two last cases it is
 less: 0.998865 and 0.996851 correspondingly.

The $\beta$ indices for different values $\alpha$ are also presented in TABLE 1.

We have also studied the extreme case when $\alpha=1$. In this situation,
after some transient period of time a state of the system become
statistically stable and character of its structure does not change. In
so
doing, we have paid attention to the following characteristics: the
relaxation
time (the time of structure formation) {\it t-rel}, the wiring of the
system {\it w}, the maximal size of a cluster {\it max-s}. 

It was revealed that {\it t-rel} and {\it max-s} depend on the system
size, but {\it w} does not.
As {\it max-s} increases with N and taking into account that in
this case the cluster distribution is better approximated by the power law
with exponent equal to -2.8 (accuracy 96\%) than by the exponential
law with
exponent equal -0.54 (accuracy 95\%) we
can suppose that in this quenched case our model is more similar to
percolation tasks. An infinite cluster would have an
infinite
size. Let us note that exponent of the critical cluster size distribution
in our model is greater (in absolute value) than one in the percolation
models in three dimention space according to \cite{Stf}, where it is
equal
-2.2, but it has the same order.

The set of these values for different system size is presented in TABLE 2.

The wiring value of the system for $\alpha=1$ is equal to 0.691.

Since
{\it w} does not depend on the system size it should be the same in
infinite systems, thus it can be envisaged as the percolation threshold in
the quenched variant of our model.
Comparing the obtained result with ones represented in \cite{perc} one can
see that ours is fairly close to percolation threshold for honeycomb lattice,
but it is rather a contingency. According to our rule of structure
construction, it can be supposed
a getting structure to be the variant of
Bethe lattice, but in our case according to the formulae $p_c=1/q$ (where
$q$ is number of links going from a node, $p_c$ -- percolation threshold, 
in our case it corresponds to
{\it w}), $q \approx 1.447$, i.e. far away from any integer number, so one
can
conclude that the created structure of our system is a tree graph with no equal
number of successors for every node. 
(According to the definition of the
Bethe lattice \cite{Efr} the number of successors for every node should
be the same).

Because of we investigate not only the quenched variant but the
possibilities
both appearance and destroying of links, besides characteristics of arising
structures, large interest has dynamics of the structure formation.
Let us note that besides characteristics of arising structures, we can
investigate dynamics of the structure formation.
Namely, we are interested in
such characteristics as the duration of deviation of the
fraction of free
elements in the system from the average.

After some transient time, the system comes to the stationary
regime.  Thus statistical division into two fractions, namely, the
fraction of free
elements and the fraction of the elements having at least a single link
takes place.
We count the average fraction of free elements in 
the system:
$\lambda=\frac{1}{NT} \sum_{i=1}^T n_i$ where {\it T} -- the total time of
counting; ${\it n_i}$ -- the number of free elements at the moment {\it
i}. Since each state of the system is not
stable, i.e., there permanently occur processes of appearing and
destroying of links, it is mirrored by the change of the fraction of free
elements
in the system ({\it dl}) (deviation from $\lambda$). 
Namely, when a link appears then {\it dl} decreases, if it vanishes
-- {\it dl} increases. 
It seems to be obvious that
deviations of different duration can be observed: both increase and
decrease of {\it
dl}. Such
deviation one can think of as the avalanches in models of self-organized
criticality \cite{BTW, mean4bsm} or as a relaxation time of the system into the
quasistationary state analogously to fluctuation motions (spin
fluctuations, order parameter fluctuations) in the phase transitions and
the critical phenomena theory \cite{Lan}. We investigated the structure
destruction processes, i.e. the probability of the {\it dl} deviation
from the average to larger values ($dl_{up}$) as a function of its
duration ($t$). In so
doing, moments $t=n/N>\lambda$ and $t=n/N<\lambda$ are considered as the
start and the end of an event correspondingly.

As it was shown by computer simulations, this characteristic is
well approximated by the power
law
$$dl_{up} \sim t^\gamma$$
with the index $\gamma=1.421 \pm 0.005$. In our opinion, it is
a remarkable result that for $t>3$ 
the curves are very close despite very different values of $\alpha$: from
1/4 to 3/4. That is also true for $t>15$ and $\alpha$=0.95, and for
$t>50$ and $\alpha=0.99$, i.e. the exponent (critical
index) does not
depend on the system parameter $\alpha$. 
Thus we have the parameter independent critical dynamics in the system. 
(Fig. 4)

In TABLE 3 one can see indexes $\gamma$ for different value of $\alpha$.

As for cluster distributions, the best correspondence is achieved for
$\alpha =1/2$
As the deviation from this value increases, for small duration
one can observe a bending of the curves to the bottom. In general, the
accuracy of the approximation was near 99.7\%.

We have also considered another case, i.e. processes of arising a
structure.
In this situation the results are turned out to be analogous to the
former case. So the picture is symmetrical.

Apart from it we calculated distribution of deviation mass. By this term
we denote deviation from the average measured in the number of elements
become free or connected during an event. Analogously to the
phase
transitions and
critical phenomena theory it can be envisage as size of fluctuation.
 The results one can see in
(Fig. 5)

Let us stress that this result is not the only one possible. Indeed the
average fraction of free elements in the system we calculated as it
mentioned above as $\lambda=\frac{1}{NT} \sum_{i=1}^T n_i$, so  it can
happen that the deviation can be shorter in time and deeper in the number
of elements in one direction than in another one with the same value of
average.

\section{Conclusion}

It seems that in spite of the simplicity of its formulation, the model
has a number of nontrivial characteristics.

After a transient period of time, a certain statistical mobile
interaction
structure is built up in the system. 
In the case of the quenched model ($\alpha =1$) the
distribution of cluster size is better approximated by the power law than 
by the exponential one, as it occurs for
percolation models. Here the critical structure kind of tree graph with no
equal number of successers is formed.

Due to very good correspondence of the distribution of the system wiring
to the Gaussian law and taking into account that the deviation is quite
small one can conclude that large fluctuations in the system occur rarely
and they rapidly vanish.

The critical character of the system dynamics does not depend on
introduced parameter $\alpha$, as it follows from distributions of time
duration of deviation from the average {\it dl}. The
distributions are
turned to be
power functions with the same index $\gamma=1.421 \pm 0.005$, so this 
model can be putted in the class SOC systems. It
looks like the principle of structure formation employed in the model
leads to a certain class of a universality.

Let us stress that in the general case while system dynamics has turned
out to be critical,
geometry of forming structures does not have such properties, as it is
shown by the exponential drop in the distribution of clusters. In addition,
in this case the index depends from $\alpha$.

In this work we didn't attempt to create the model of SOC system as it
done in \cite{Fl}, but we revealed such properties in the simplest model
of structure formation. Maybe this work can be considerable as an approach
to understanding of the mechanisms of self-organization to criticality.

{\small Acknowlegement.
The author greatly acknowledges Konstantin Mardanov for the help during
a preparation of the manuscript.}


\newpage
\centerline{FIGURES}

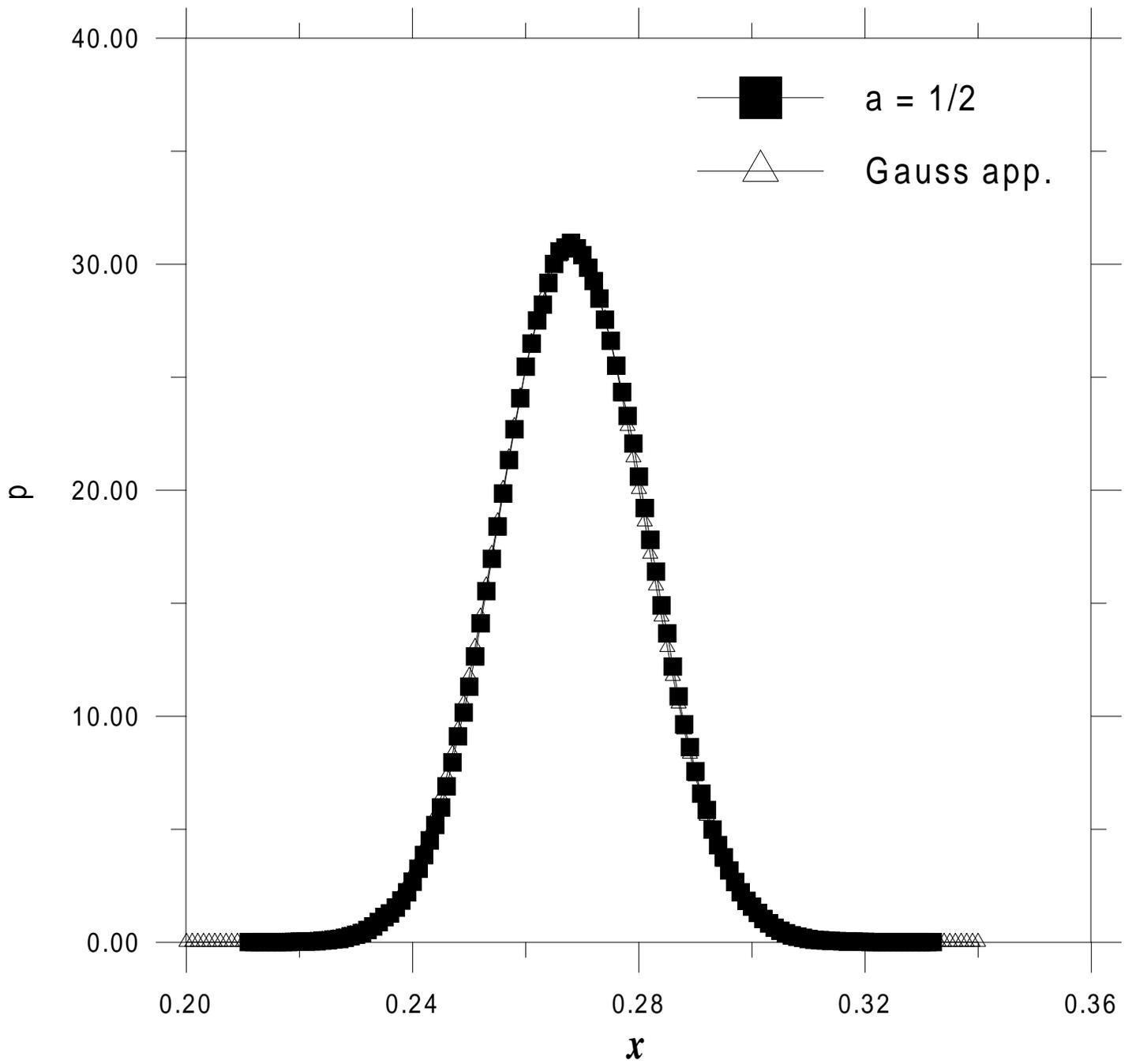
FIG. 1. Probability distribution of the system wiring p(x) for parameter
$\alpha=1/2$ (squares)  in comparison with Gaussian distribution with the
same parameters (triangles). Size of the system N=1000.

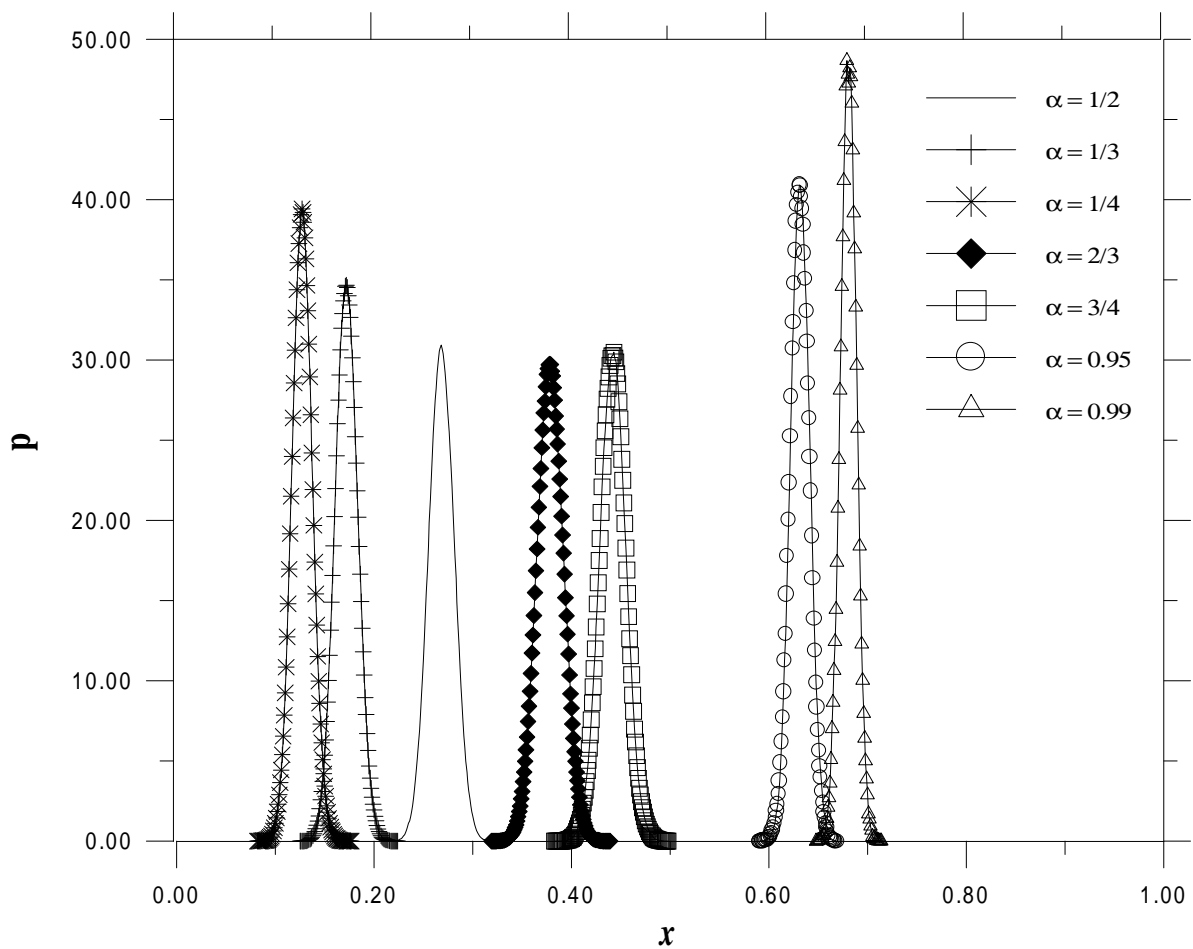
FIG. 2. Probability distribution of the system wiring p(x) for parameters:
$\alpha=1/2$ (line only); $\alpha=1/4$ (stars); $\alpha=1/3$ (cross);
$\alpha=2/3$ (diamonds); $\alpha=3/4$ (squares); $\alpha=0.95$ (circles);
$\alpha=0.99$ (triangles).
Size of the system N=1000.

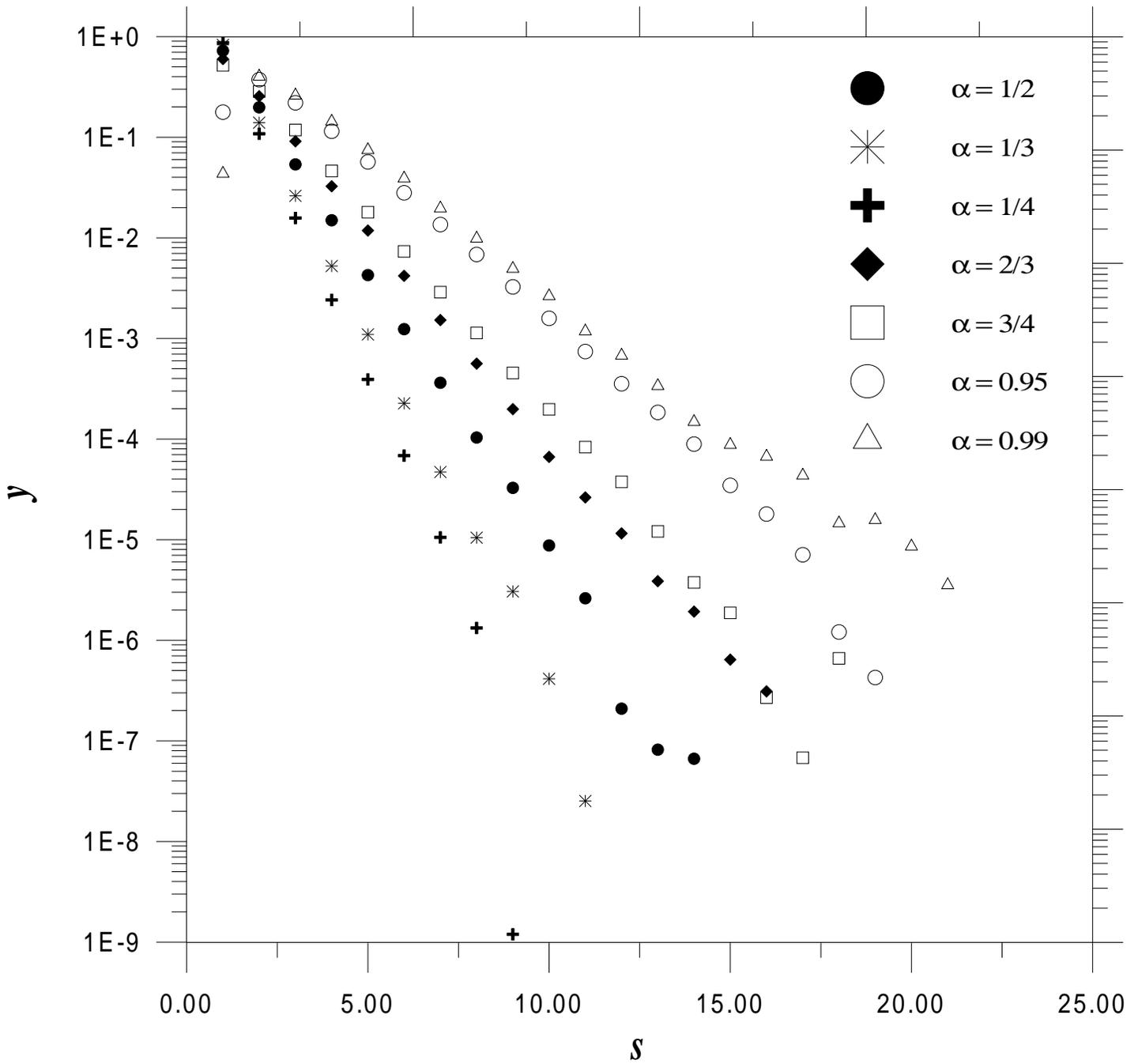
FIG. 3. Distribution of clusters size y(s) for parameters:
$\alpha=1/2$ (dots); $\alpha=1/4$ (stars); $\alpha=1/3$ (cross);
$\alpha=2/3$ (diamonds); $\alpha=3/4$ (squares); $\alpha=0.95$ (circles);
$\alpha=0.99$ (triangles).
Size of the system N=1000.

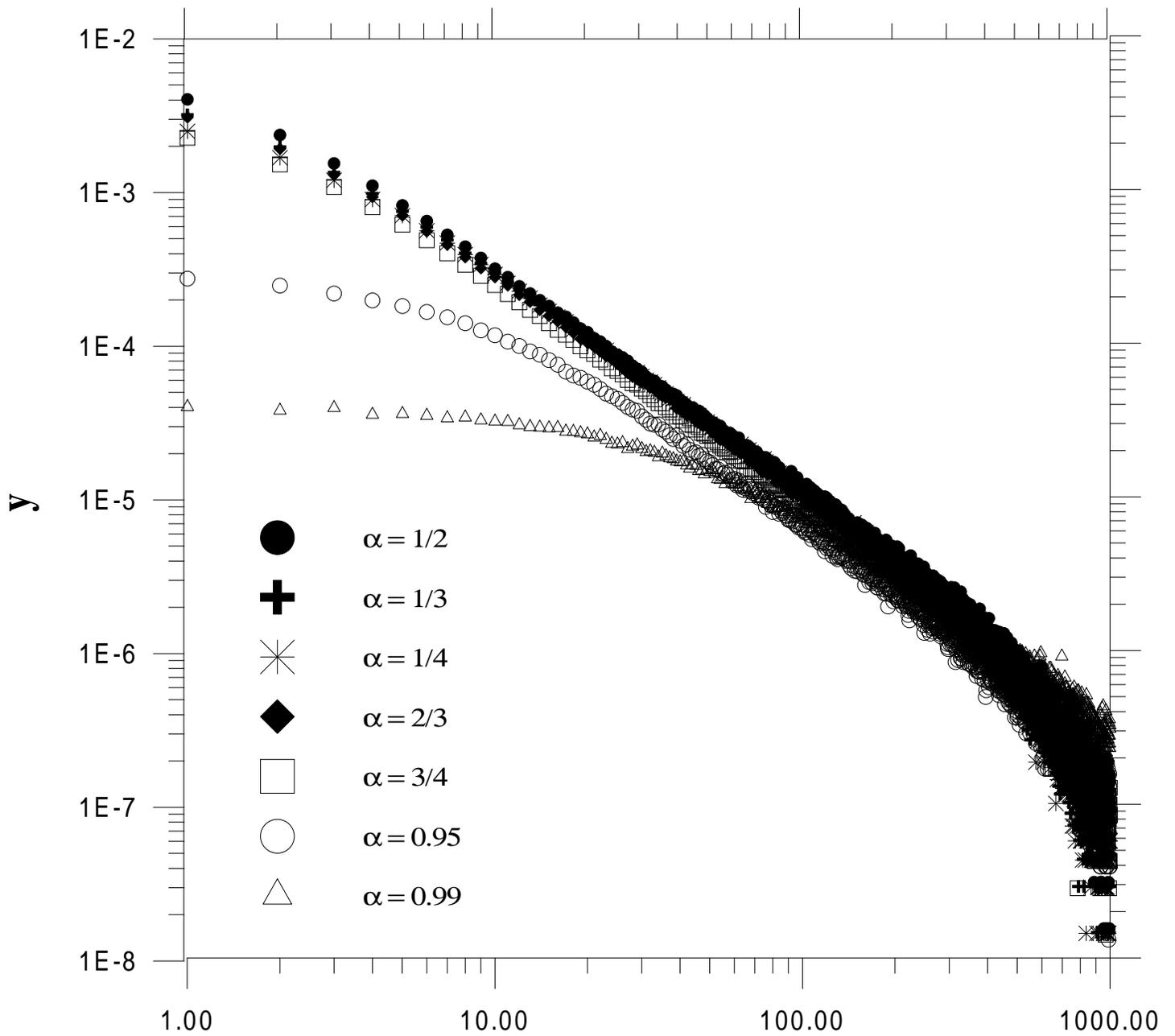
FIG. 4. Distribution of deviation from average of free elements fraction in
the system y(x) for parameters:
$\alpha=1/2$ (dots); $\alpha=1/4$ (stars); $\alpha=1/3$ (cross);
$\alpha=2/3$ (diamonds); $\alpha=3/4$ (squares); $\alpha=0.95$ (circles);
$\alpha=0.99$ (triangles).
Size of the system N=500.

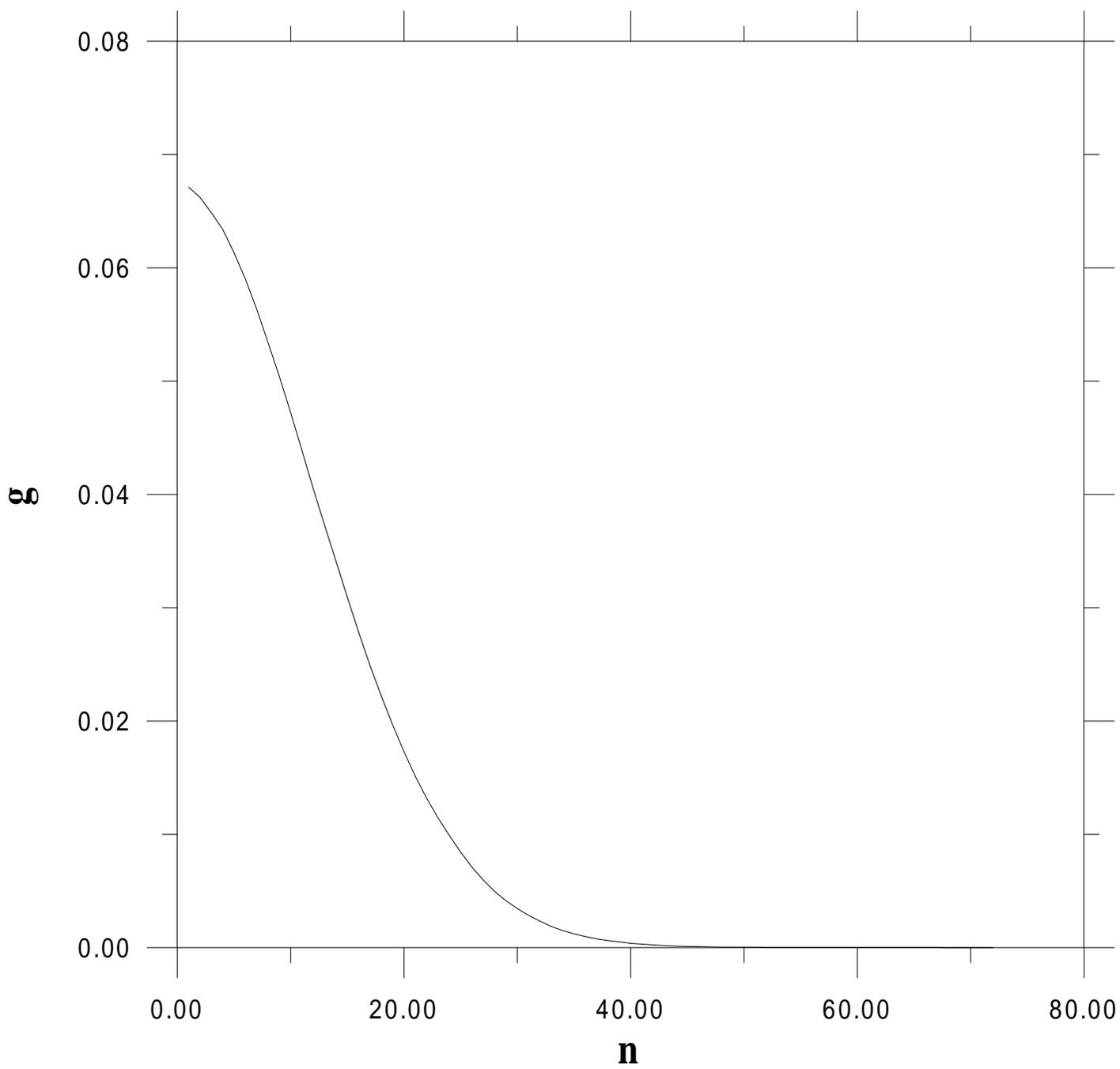
FIG. 5. Distribution of deep of deviation from average of free elements
fraction in the system g(n) for $\alpha=3/4$.
Size of the system N=500.

\newpage
\centerline{TABLES}

\begin{center}

\begin{tabular}{|c|c|c|c|c|c|c|}
\hline
\tt $\alpha$ & dl & a & $\sigma$ & $\beta$ & max-s & range
\\
\hline
\tt 1/4 & 0.764 & 0.127 & 0.01 & -1.88235 & 9 &[0.082;0.177]
\\
\hline
\tt 1/3 & 0.689 & 0.172 & 0.0115 & -1.60361 & 11 &[0.125;0.224]
\\
\hline
\tt 1/2 & 0.536 & 0.268 & 0.013 & -1.24915 & 14 &[0.211;0.332]
\\
\hline
\tt 2/3 & 0.378 & 0.378 & 0.0135 & -0.98363 & 16 &[0.319;0.439]
\\
\hline
\tt 3/4 & 0.295 & 0.442 & 0.013 & -0.909985 & 18 &[0.383;0.498]
\\
\hline
\tt 0.95 & 0.066 & 0.632 & 0.0095 & -0.725898 & 19 &[0.590;0.669]
\\
\hline
\tt 0.99 & 0.014 & 0.681 & 0.008 & -0.645233 & 21 &[0.648;0.714]
\\
\hline
\end{tabular}
\end{center}

TABLE 1

In this table values characterized of the properties of a forming structure
for different parameter $\alpha$ values are presented. Where

$\alpha$ is the probability of link appearance

{\it dl} - fraction of free elements in the system

{\it a} - average value of the wiring

$\sigma$ - deviation of the wiring

$\beta$ - index of the cluster size distribution

{\it max-s} - maximal size of a cluster

{\it range} - range of non-zero values of the wiring

We investigated the systems of 1000 elements (N=1000).

\begin{center}

\begin{tabular}{|c|c|c|}
\hline
\tt N & t-rel &  max-s
\\
\hline
\tt 100 & 446 & 6
\\
\hline
\tt 200 & 1405 & 9
\\
\hline
\tt 300 & 1124 & 11
\\
\hline
\tt 400 & 2697 & 9
\\
\hline
\tt 500 & 2502 & 12
\\
\hline
\tt 600 & 2930 & 10
\\
\hline
\tt 700 & 3969 & 8
\\
\hline
\tt 800 & 4766 & 13
\\
\hline
\tt 900 & 5700 & 10
\\
\hline
\tt 1000 & 7012 & 10
\\
\hline
\tt 1500 & 7675 & 12
\\
\hline
\tt 2000 & 15367 & 13
\\
\hline
\tt 2500 & 20632 & 11
\\
\hline
\tt 3000 & 21785 & 11
\\
\hline
\tt 3500 & 31820 & 16
\\
\hline
\end{tabular}
\end{center}

TABLE 2

In this table the relaxation time (the time of structure construction)
{\it t-rel} and the maximal size of a cluster {\it max-s} for different
system size (N) in border case ($\alpha=1$) are presented.

\begin{center}

\begin{tabular}{|c|c|}
\hline
\tt $\alpha$ & $\gamma$
\\
\hline
\tt 1/4  & -1.42006
\\
\hline
\tt 1/3  & -1.42167
\\
\hline
\tt 1/2  & -1.42349
\\
\hline
\tt 2/3  & -1.41858
\\
\hline
\tt 3/4  & -1.42415
\\
\hline
\tt 0.95 & -1.4161
\\
\hline
\tt 0.99 & -1.42407
\\
\hline
\end{tabular}
\end{center}

TABLE 3

In this table one can see critical indexes $\gamma$ for different values of
$\alpha$. N=500.

\end{document}